\documentclass[reprint,amsmath,amssymb,aps,superscriptaddress,pra]{revtex4-2}
\usepackage{graphicx}
\usepackage{bm}
\usepackage{tikz}
\usepackage{physics}
\usetikzlibrary{quantikz2}

\begin{document}
  \title{Entanglement-assisted phase estimation algorithm for calculating dynamical response functions}
  \author{Rei Sakuma}
  \affiliation{Materials Informatics Initiative, RD Technology \& Digital Transformation Center, JSR Corporation,
    3-103-9 Tonomachi, Kawasaki-ku, Kawasaki, 210-0821, Japan}
  \affiliation{Quantum Computing Center, Keio University, 3-14-1 Hiyoshi, Kohoku-ku, Yokohama 223-8522, Japan}

  \author{Shu Kanno}
  \affiliation{Mitsubishi Chemical Corporation, Science \& Innovation Center, Yokohama, 227-8502, Japan}
  \affiliation{Quantum Computing Center, Keio University, 3-14-1 Hiyoshi, Kohoku-ku, Yokohama 223-8522, Japan}

  \author{Kenji Sugisaki}
  \affiliation{Quantum Computing Center, Keio University, 3-14-1 Hiyoshi, Kohoku-ku, Yokohama 223-8522, Japan}
  \affiliation{Graduate School of Science and Technology, Keio University, 7-1 Shinkawasaki, Saiwai-ku, Kawasaki, Kanagawa 212-0032, Japan}
  \affiliation{Centre for Quantum Engineering, Research and Education, TCG Centres for Research and Education in Science and Technology, Sector V, Salt Lake, Kolkata 700091, India}

  \author{Takashi Abe}
  \affiliation{Quantum Computing Center, Keio University, 3-14-1 Hiyoshi, Kohoku-ku, Yokohama 223-8522, Japan}
  \author{Naoki Yamamoto}
  \affiliation{Quantum Computing Center, Keio University, 3-14-1 Hiyoshi, Kohoku-ku, Yokohama 223-8522, Japan}
  \affiliation{Department of Applied Physics and Physico-Informatics, Keio University, 3-14-1 Hiyoshi, Kohoku-ku, Yokohama 223-8522, Japan}

  \date{\today}
  \begin{abstract}
    Dynamical response functions are fundamental quantities
    to describe the excited-state properties in quantum many-body systems. Quantum algorithms have been proposed to evaluate these quantities by means of quantum phase estimation (QPE), where the energy spectra are directly extracted from the QPE measurement outcomes in the frequency domain. 
    Accurate estimation of excitation energies and transition probabilities with these QPE-based approaches is, however, challenging because of the problem of spectral leakage (or peak broadening) which is inherent in the QPE algorithm. 
    To overcome this issue, in this work we consider an extension of the QPE-based approach adopting the optimal entangled input states, which is known to achieve the Heisenberg-limited scaling for the estimation precision. 
    We show that with this method the peaks in the calculated energy spectra are more localized than those calculated by the original QPE-based approaches,
    suggesting the mitigation of the spectral leakage problem.
    By analyzing the probability distribution with the entangled phase estimation, we propose a simple scheme to better estimate both the transition energies and the corresponding transition probabilities of the peaks of interest in the spectra. The validity of our prescription is demonstrated by
    numerical simulations in various quantum many-body problems: the spectral function of a simple electron-plasmon model in condensed-matter physics,
    the dipole transitions of the H$_2$O molecule in quantum chemistry, and the electromagnetic transitions of the $^6$Li nucleus in nuclear physics.
  \end{abstract}
  \maketitle

  \section{introduction}

  Studying the excited-state properties of many-body quantum systems are not only of academic interest but also crucial for industrial applications. 
  Dynamical response functions and correlation functions
  are fundamental quantities that reflect the excited-state structures of a system and that describe responses of the system to an external field, which 
  are directly connected to
  experimental results.
  The calculation of the response functions of many-body systems on a classical computer, however,
  remains a theoretical challenge
  as the computational cost scales exponentially with respect to the system size.

  Quantum simulation of many-body systems is considered as a promising application for a quantum computer, and
  a number of papers have investigated how to compute the fermionic
  Green function~\cite{PhysRevX.6.031045,PhysRevA.93.032303,PhysRevResearch.2.033281,PhysRevA.101.012330,PhysRevA.104.032422,keen2021quantum,
    PhysRevA.103.032404,jamet2022quantum,Keen2022hybridquantum,PhysRevResearch.4.043011,
    PhysRevResearch.4.043038,doi:10.1021/acs.jctc.3c00150,PhysRevA.108.012618,ralli2023calculating,
    greenediniz2023quantum,kokcu2023linear,kowalski2024capturing}
  as well as other dynamical response functions~\cite{kokcu2023linear,PhysRevLett.113.020505,PhysRevA.92.062318,PhysRevC.100.034610,PhysRevResearch.2.033043,PhysRevResearch.2.033043,
    PhysRevResearch.2.033324,PhysRevLett.129.240501,
    doi:10.1021/acs.jctc.3c00731} efficiently on a quantum computer.
  A standard algorithm to compute response functions
  is to perform the real-time propagation of a given system
  on a quantum computer, followed by the Fourier
  transformation on a classical computer to calculate the energy spectra.
  This hybrid time-domain approach requires repeated measurements at each point on the time grid, and sometimes
  it is combined with further classical postprocessing, such as the convolution with some damping function~\cite{kokcu2023linear}
  or interpolation/extrapolation of the results.
  These postprocessing operations make it difficult to extract
  from the calculated energy spectra
  the exact values of the transition energies and the corresponding transition probabilities,
  which are two important excited-state properties of the system.

  Quantum phase estimation (QPE)~\cite{kitaev1995quantum,PhysRevLett.83.5162} is one of the most
  fundamental quantum algorithms, and
  approaches have been proposed to use QPE in the calculation of
  dynamical response functions~\cite{PhysRevA.92.062318,PhysRevC.100.034610,PhysRevA.101.012330,PhysRevResearch.2.033043,kowalski2024capturing}.
  These methods may be considered as a time-domain approach that performs both the real-time propagation and the
  Fourier transformation on a quantum computer.
  The energy spectra are obtained directly by collecting the QPE measurement results in the frequency domain, and
  the resolution of the spectra can systematically be improved by increasing the number of qubits.
  These approaches, therefore, offer a promising way to calculate excited-state properties of a system
  on a quantum computer,
  combined with a number of ground-state preparation algorithms
  proposed~\cite{Peruzzo2014,D1CS00932J,PhysRevLett.102.130503,10.1063/1.5027484,Lin2020nearoptimalground,PRXQuantum.3.040305,
    kerzner2023squareroot,ding2023singleancilla}.

  Although the QPE-based approaches are attractive, QPE suffers a known problem of ``spectral leakage''~\cite{9762511};
  if the phase is not exactly on the QPE frequency grid, the measurement results ``leak'' from the exact position, resulting in
  a long tail in the measured probability distribution.
  This hinders the accurate estimation of the transition energies and probabilities from the spectra
  calculated with the QPE-based approaches with a finite number of qubits.
  Several approaches have been proposed to address
  this leakage problem in the general context of QPE~\cite{9762511,10.5555/2012098.2012106,PhysRevLett.117.010503},
  but to the best of our knowledge, no previous study has focused on the spectral properties of physical systems.
  Since peak broadening in calculations can be a major obstacle in making comparisons with experiments,
  addressing the leakage problem in the calculation of energy spectra is an urgent issue.
  
  In this work, we consider mitigating this problem by employing entangled phase estimation (EPE), which
  performs QPE with the optimal entangled input state. 
  Studies have shown~\cite{PhysRevA.54.4564,PhysRevLett.82.2207,PhysRevLett.98.090501,4655455,PhysRevX.8.041015} that with the optimal entangled input state 
  the mean squared error (MSE) of the estimated phase scales as $\mathcal{O}(N^{-2})$
  with $N$ the number of unitary operations for QPE,
  which is known as the Heisenberg limited scaling. 
  We show that the probability distribution obtained with EPE has characteristics different from the original QPE-based one; 
  most notably, the calculated spectra in EPE have much more localized peaks, thanks to the entangled input. 
  Through the analysis of the EPE-based probability distribution,
  we propose a simple scheme to estimate the transition energies and corresponding transition probabilities
  of specified peaks in the spectra.
  We show that when this scheme is combined with the EPE-based calculation,
  the estimation error of the transition probabilities for isolated peaks is less than 1\%
  in the limit of infinite sampling. This is a significant improvement over the original QPE-based calculation,
  whose worst-case error in the same setting is $\approx 10$\%.
  To demonstrate the usefulness of the proposed approach,
  we present numerical simulations of the spectral function of an electron-plasmon system,
  dipole transitions in quantum chemistry, and transition probabilities in nuclear physics.

  \section{method}
  \subsection{General formalism}
  \begin{figure*}
    \includegraphics{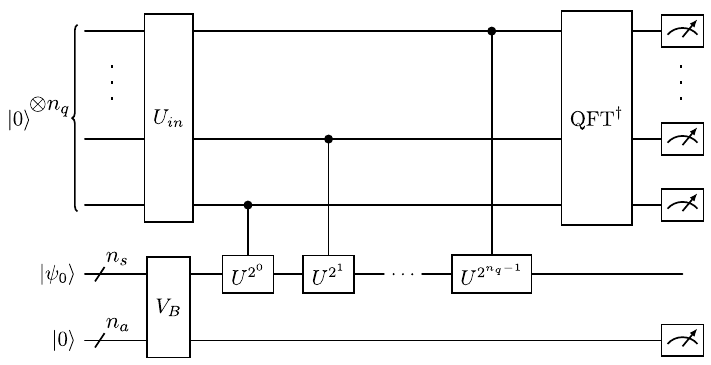}
    \caption{The quantum circuit to evaluate $S_{B}(\omega)$ ($\bar{S}_{B}(\omega_{k})$ in Eq.~\eqref{eq:sbar_b}).
    Here $U=\exp[i (H - E_{0})T]$ and $\mathrm{QFT^\dagger}$ represents the inverse quantum Fourier transform.}
    \label{fig:qc_sbw}
  \end{figure*}
  We first outline the QPE-based approaches proposed
  in Refs.~\cite{PhysRevA.92.062318,PhysRevC.100.034610,PhysRevA.101.012330,PhysRevResearch.2.033043,kowalski2024capturing}.
  For a system described by the Hamiltonian $H$,
  our interest in this work is to calculate for a given operator $B$ the following
  dynamical response function \cite{PhysRevC.100.034610}
  \begin{equation}
    S_{B}(\omega) = \sum_{s} |\bra{\psi_{s}} B \ket{\psi_{0}}|^{2} \delta (\omega - \Delta E_{s}),
    \label{eq:sbw}
  \end{equation}
  where $s$ label the eigenstates of the Hamiltonian with $s=0$ corresponding to the ground state,
  $\ket{\psi_{s}}$ are the eigenstates with associated eigenenergies $E_{s}$, and $\Delta E_{s} = E_{s} - E_{0}$ are the transition energies.
  The prefactors or weights of the $\delta$ functions in Eq.~\eqref{eq:sbw},
  $|\bra{\psi_{s}} B \ket{\psi_{0}}|^{2}$, are the transition probabilities.
  Equation \eqref{eq:sbw} is related, via the Fourier transformation, to the Lehmann representation of
  the autocorrelation function $C_{BB}(t)$, where
  \begin{equation}
    C_{B_{1}B_{2}}(t) = \bra{\psi_{0}} e^{iHt} B_{1} e^{-iHt} B_{2} \ket{\psi_{0}}
    \label{eq:corr}
  \end{equation}
  is the correlation function. We set $\hbar = 1$ throughout this work.
  
  The general form of the quantum circuit used to evaluate Eq.~\eqref{eq:sbw} via QPE
  is shown in Fig.~\ref{fig:qc_sbw}, which consists of three parts: The first register is for the QPE algorithm with $n_{q}$ qubits,
  and $U_{in} = \textrm{H}^{\otimes n_{q}}$ creates the unentangled input state of QPE with $\textrm{H}$ the
  Hadamard gate.
  The second register is initialized to  the ground state of the system $\ket{\psi_{0}}$ described with $n_{s}$ qubits,
  and $U=\exp [i (H - E_{0}) T]$ is the time propagation operator.
  Here $T$ is an artificial time introduced to scale the energy
  so that the energy range of interest becomes $[0, \frac{2 \pi}{T}]$.
  The third (optional) register is an ancilla register with $n_{a}$ qubits, which may be used to
  encode the $B$ operator through the gate $V_{B}$. The measurement outcomes of the third register are used
  for postselection.
  As $B \ket{\psi_{0}}$ is expanded as $B \ket{\psi_{0}} = \sum_{s} \bra{\psi_{s}} B \ket{\psi_{0}} \ket{\psi_{s}}$,
  the measurement results of the QPE register
  have a peak  for each $s$ near $\Delta E_s=E_{s} - E_{0}$ with the weight proportional to
  $|\bra{\psi_{s}} B \ket{\psi_{0}}|^{2}$. 
  More precisely, with a $V_{B}$-dependent normalization factor $\mathcal{N}$,
  the probabilities of
  measuring $k=0,1,2,\dots,2^{n_{q}}-1$ in the QPE register are given as
  \begin{equation}
    \mathcal{P}(k) = \mathcal{N} \sum_{s} |\bra{\psi_{s}} B \ket{\psi_{0}}|^{2} P(k|\Delta E_{s} T), \label{eq:p_k}
  \end{equation}
  where
  \begin{eqnarray}
    P(k|\theta) &=& \frac{1}{N_{q}}
    \Bigl|\sum_{j=0}^{N_{q}-1} a_{j} e^{i \bigl(\theta - \frac{2 \pi k}{N_{q}}\bigr)j}\Bigr|^{2} \nonumber\\
    &=& \frac{1}{N_{q}^{2}}
    \frac{\sin^{2}\frac{\theta N_{q}}{2}}
    {\sin^{2}\frac{1}{2}(\theta - \frac{2 \pi k}{N_{q}})},
    \label{eq:p_s_k} \\
    a_{j}&=&\frac{1}{\sqrt{N_{q}}}.
    \label{eq:a_j_qpe}
  \end{eqnarray}
  Here  $N_{q}=2^{n_{q}}$ and $a_{j}$ are the coefficients of the input state for QPE
  \begin{equation}
    \ket{\phi^{\textrm{QPE}}_{in}} = \sum_{j=0}^{N_{q}-1}a_{j}\ket{j}
    \label{eq:qpe_in}
  \end{equation}
  created from $U_{in} = \textrm{H}^{\otimes n_{q}}$ in Fig.~\ref{fig:qc_sbw}.
  From $\mathcal{P}(k)$, an approximation of $S_{B}(\omega)$ is obtained as
  \begin{equation}
    \bar{S}_{B}(\omega_{k}) = \frac{1}{\mathcal{N}}\mathcal{P}(k),
    \label{eq:sbar_b}
  \end{equation}
  where $\omega_{k} = \frac{2 \pi}{N_{q} T} k$ 
  are the frequency grid points in QPE.

  In this approach one can also evaluate off-diagonal quantities, namely,
  the Lehmann representation of
  $C_{B_{1}B_{2}}(t)$ with $B_{1}\neq B_{2}$, by considering
  $B_{\pm} = B_{1} \pm B_{2}$ and $B_{i\pm} = B_{1} \pm i B_{2}$.
  In the construction of the quantum circuit in Fig.~\ref{fig:qc_sbw}
  it is assumed that the ground state $\ket{\psi_{0}}$ and the ground-state energy $E_{0}$ are known a priori,
  and that one has access to controlled $U$ gates.
  One possible way to avoid the dependencies on $E_{0}$ and controlled $U$ gates
  using the concept of quantum phase difference estimation~\cite{BPDE_2021, QPDE_2023} is discussed in Appendix.

\begin{figure}
    \includegraphics{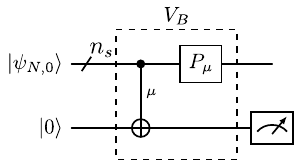}
    \caption{The $V_{B}$ gate in Fig.~\ref{fig:qc_sbw} for evaluating $A^{\textrm{particle}}_{\mu}(\omega)$ and $A^{\textrm{hole}}_{\mu}(\omega)$.
    Here $P_{\mu} = \Bigl[ \prod_{\nu<\mu} Z_{\nu}\Bigr] X_{\mu}$ and the
    CNOT gate takes the $\mu$-th qubit in $\ket{\psi_{N,0}}$ as control.}
    \label{fig:qc_ajw}
  \end{figure}

  One important application of this approach considered in Refs.~\cite{PhysRevA.101.012330,kowalski2024capturing}
  is the calculation of spectral function $A(\omega)$ in the electron many-body problem. This case corresponds to
  $B = c^{\dagger}_{\mu}$ and $c_{\mu}$, which are the electron
  creation and annihilation operators for one-particle states $\mu$, respectively.
  The spectral function of an $N$-electron system can be decomposed into
  the electron-creation (particle) part and the electron-annihilation (hole) part as
  \begin{equation}
    A(\omega) = \sum_{\mu} \bigl( A_{\mu}^{\mathrm{particle}}(\omega) + A_{\mu}^{\mathrm{hole}}(\omega) \bigr),
  \end{equation}
  where
  \begin{eqnarray}
    A_{\mu}^{\mathrm{particle}}(\omega) &=&
    \sum_{s} |\bra{\psi_{N+1,s}} c^{\dagger}_{\mu} \ket{\psi_{N,0}} |^{2} \nonumber\\
    && \qquad \times \delta(\omega - (E_{N+1,s} - E_{N,0})), \label{eq:aw_particle}\\
    A_{\mu}^{\mathrm{hole}}(\omega) &=&
    \sum_{s'} |\bra{\psi_{N-1,s'}} c_{\mu} \ket{\psi_{N,0}} |^{2} \nonumber\\
    && \qquad \times \delta(\omega - (E_{N,0} - E_{N-1,s'})).
    \label{eq:aw_hole}
  \end{eqnarray}
  Here $\ket{\psi_{N,0}}$ is the ground state of the $N$-electron system with energy $E_{N,0}$, and
  $\ket{\psi_{N\pm 1,s}}$ are the eigenstates of the $(N\pm 1)$-electron systems
  with corresponding eigenenergies $E_{N\pm 1,s}$.
  The spectral function $A(\omega)$ is related to
  the diagonal elements of the retarded one-particle Green function at zero temperature in the frequency domain~\cite{fetter}
  and can be compared with experimental spectra measured with photoemission and inverse photoemission spectroscopies.
  When the Jordan--Wigner transformation (JWT) is used,
  the $V_{B}$ gate in Fig.~\ref{fig:qc_sbw} has a particularly simple form requiring only one ancilla qubit;
  in this case,
  $c_{\mu}$ and $c^{\dagger}_{\mu}$ are expanded with Pauli operators as
  \begin{eqnarray}
    c_{\mu} &=& \Bigl(\prod_{\nu<\mu} Z_{\nu} \Bigr) \frac{X_{\mu}+iY_{\mu}}{2}, \\
    c^{\dagger}_{\mu} &=& \Bigl(\prod_{\nu<\mu} Z_{\nu} \Bigr) \frac{X_{\mu}-iY_{\mu}}{2},
  \end{eqnarray}
  and the $V_{B}$ gate to calculate $A_{\mu}^{\textrm{particle}}(\omega)$ and $A_{\mu}^{\textrm{hole}}(\omega)$
  is shown in Fig.~\ref{fig:qc_ajw}.
  As the ancilla qubit at the bottom becomes $1 (0)$ when the $\mu$-th orbital is occupied (unoccupied),
  by measuring this qubit as well as the QPE qubits both the particle and the hole parts of the spectral function
  can be sampled simultaneously.
  The true $A^{\textrm{particle}}_{\mu}(\omega)$ and $A^{\textrm{hole}}_{\mu}(\omega)$
  should satisfy
  \begin{equation}
    \int_{-\infty}^{+\infty} \Bigl[
    A^{\textrm{particle}}_{\mu}(\omega) + A^{\textrm{hole}}_{\mu}(\omega) \Bigr]
    d\omega = 1,
  \end{equation}
  as can be verified by considering the anticommutation relation $c_{\mu}c^{\dagger}_{\mu} + c^{\dagger}_{\mu}c_{\mu}=1$.
  This corresponds to setting $\mathcal{N} = 1$ in Eq.~\eqref{eq:sbar_b}.

  For a more general $B$, several approaches can be used to block-encode $B$.
  The linear combination of unitary (LCU) approach~\cite{childs2012} can be employed
  when $B$ is represented as a sum of unitary operators as
  $B = \sum_{l=1}^{N_{B}} \lambda_{l} U_{l}$, where $\lambda_{l} \ge 0$ and $U_{l}$ are unitary operators
  that act on the system qubits.
  In this case, $V_{B}$ is expressed with $n_{a} = \lceil \log_{2}N_{B} \rceil$ ancilla qubits
  as $ V_{B} = \textrm{PREP}^{\dagger} \cdot \textrm{SELECT} \cdot \textrm{PREP}$, where
  \begin{eqnarray}
    \textrm{PREP} \ket{0}^{\otimes n_{a}} &=& \frac{1}{\sqrt{\sum_{l}\lambda_{l}}}\sum_{l} \sqrt{\lambda_{l}} \ket{l}, \\
    \textrm{SELECT} &=& \sum_{l} U_{l} \otimes \ket{l}\bra{l},
  \end{eqnarray}
  and $\bar{S}_{B}(\omega_{k})$ is obtained by multiplying the probabilities of measuring $\ket{k} \otimes I^{\otimes n_{s}} \otimes \ket{0}^{\otimes n_{a}}$
  states in Fig.~\ref{fig:qc_sbw} by $\bigl(\sum_{l} \lambda_{l}\bigr)^{2}$.

  \subsection{Use of entangled phase estimation}
\begin{figure}
    \includegraphics{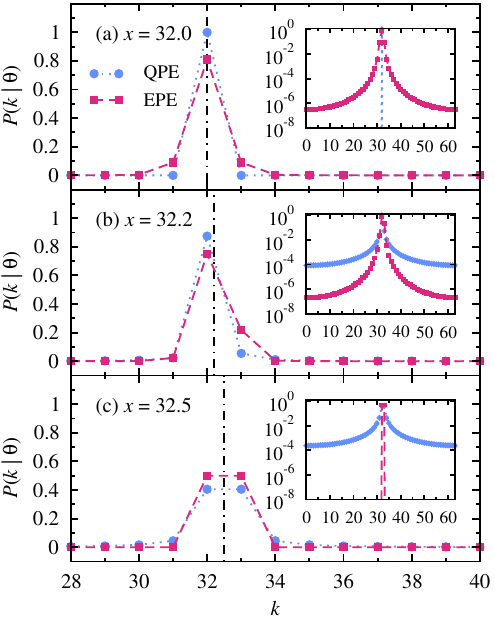}
    \caption{Comparison of $P(k|\theta)$ in QPE
      and EPE for $n_{q}=6$ ($N_{q}=64$).
      The positions of $\theta$ are indicated by the vertical dash-dotted lines, and
      $x$ is defined as $\theta = \frac{2\pi}{N_{q}}x$. (a) $x=32.0$. (b) $x=32.2$. (c) $x=32.5$. The insets show
    $P(k|\theta)$ in logarithmic scale.}
    \label{fig:qpeopt1}
  \end{figure}

  The spectra obtained in the QPE-based approaches, $\bar{S}_{B}(\omega_{k})$,
  have nonzero values for all $\omega_{k}$ unless all
  $\Delta E_{s} T$ values are represented with $n_{q}$ bits; this
  means that the peaks in $\bar{S}_{B}(\omega_{k})$ can have long tails.
  This effect, called spectral leakage, can be mitigated by using entangled input states
  for QPE~\cite{PhysRevA.54.4564,PhysRevLett.82.2207,PhysRevLett.98.090501,4655455,PhysRevX.8.041015,10.1116/5.0147954},
  where the coefficients $a_{j}$ in Eqs.~\eqref{eq:a_j_qpe} and \eqref{eq:qpe_in}
  are replaced by optimal ones.
  In this work, we focus on the following form
  \begin{equation}
    a_{j} = \sqrt{\frac{2}{N_{q}}} \sin \frac{\pi j}{N_{q}}.
    \label{eq:a_j_qpe_opt}
  \end{equation}
  The corresponding $U_{in}$ in Fig.~\ref{fig:qc_sbw} can be implemented by using the quantum Fourier transform gate~\cite{PhysRevD.106.034503} (or a shallower gate using ancilla qubits~\cite{PhysRevX.8.041015}), and with this form $P(k|\theta)$ in Eq.~\eqref{eq:p_k} becomes
  \begin{equation}
    P(k|\theta) = \frac{\cos^{2}\frac{\theta N_{q}}{2}}{2 N^{2}_{q}}
    \frac{\sin^{2}\frac{\pi}{N_{q}}}
    {\sin^{2}\frac{\theta_{k+}}{2}
     \sin^{2}\frac{\theta_{k-}}{2}
    },
    \label{eq:p_k_epe}
  \end{equation}
  where
  \begin{equation}
    \theta_{k\pm} = \theta - \frac{2\pi k}{N_{q}} \pm \frac{\pi}{N_{q}}.
  \end{equation}
  Figure~\ref{fig:qpeopt1} compares $P(k|\theta)$ in QPE (Eq.~\eqref{eq:p_s_k}) and
  EPE (Eq.~\eqref{eq:p_k_epe}) for three values of $\theta = \frac{2 \pi}{N_{q}}x$.
  The most prominent feature of the EPE probability distribution is that, as can be seen in the inset of
  Fig.~\ref{fig:qpeopt1}(b),
  the tail probability decreases more rapidly than the QPE one for general $\theta$.
  It has been shown that because of this property
  the MSE of the estimated phase
  scales as $\mathcal{O}(\frac{1}{N^{2}_{q}})$ for large $N_{q}$~\cite{PhysRevLett.98.090501,4655455}.
  As in the case of Fig.~\ref{fig:qpeopt1}(a), when $\theta$ coincides with some $\frac{2\pi k_{0}}{N_{q}}$, where $k_0$ is an integer,
  in the original QPE algorithm $P(k|\theta)$ becomes $\delta_{k,k_{0}}$, whereas
  in EPE $P(k=k_{0}|\theta) = 2 \Bigl(N^{2}_{q}\tan^{2} \frac{\pi}{2 N_{q}}\Bigr)^{-1} < 1$. This
  means that in EPE when $\Delta E_{s}$ coincides with some $\omega_{k_{0}} = \frac{2 \pi k_{0}}{T}$,
  $\bar{S}_{B}(\omega_{k_{0}})$ is smaller than $|\bra{\psi_{s}}B\ket{\psi_{0}}|^{2}$,
  and $\bar{S}_{B}(\omega_{k})$ has nonzero tail probabilities.
  When $\theta$ is exactly at the midpoint of two grid points, as in Fig.~\ref{fig:qpeopt1}(c),
  $P(k|\theta)$ in EPE takes nonzero values
  only at the two grid points, and this is another distinct feature of EPE.

  \subsection{Estimation of transition energies $\Delta E_{s}$ and
  the transition probabilities $|\bra{\psi_{s}}B\ket{\psi_{0}}|^{2}$
  }
  \label{subsec:estimate}
  
  First recall that,
  while the exact $S_{B}(\omega)$ in Eq.~\eqref{eq:sbw} consists of $\delta$ functions, 
  $\bar{S}_{B}(\omega_{k})$ in general has nonzero values for all $\omega_{k}$.
  Consequently, the highest value in $\bar{S}_{B}(\omega_{k})$ associated with a state $s$ does not necessarily coincide with
  the true weight of the peak, $|\bra{\psi_{s}}B\ket{\psi_{0}}|^{2}$.
  In the following, we consider a simple scheme to estimate
  $\Delta E_{s}$ and $|\bra{\psi_{s}}B\ket{\psi_{0}}|^{2}$ 
  for a specified peak.

  Near $\omega_k \approx \Delta E_{s}$, the dominant contribution to
  $\bar{S}_{B}(\omega_{k})$ comes from a single state $s$, therefore
  one may neglect the contributions from all $s' \neq s$ in Eq.~\eqref{eq:p_k}.
  With this approximation, we propose the estimators for 
  $\Delta E_{s}$ and $|\bra{\psi_{s}}B\ket{\psi_{0}}|^2$ as
  \begin{eqnarray}
    \Delta E^{\textrm{est}}_{s} &=& \frac{\sum_{k}{}^{'} \omega_{k} p_{k}}{\sum_{k}{}^{'} p_{k}},
    \label{eq:est_e}\\
    |\bra{\psi_{s}}B\ket{\psi_{0}}|^{2}_{\textrm{est}} &=& \frac{1}{\mathcal{N}}\sum_{k}{}^{'} p_{k},
    \label{eq:est_h}
  \end{eqnarray}
  where the primes indicate that the sum is restricted to the points in the vicinity of the peak associated with the state $s$, and
  $p_{k}$ are the probabilities of obtaining $k$ in the measurement.
  In practice,
  for each $s$ we calculate the sum in Eqs.~\eqref{eq:est_e} and \eqref{eq:est_h}
  using $r$ highest-probability points near the observed peak, with $r$ a positive integer.
  Note that here we do not take the maximum likelihood estimator for $\Delta E_s$, i.e., the point $\omega_{k}$
  corresponding to the highest probability.
  The reason for this is that, as can be seen in Fig.~\ref{fig:qpeopt1} (b), even in ideal numerical simulations
  the maximum likelihood estimator has a bias from the true value of $\Delta E_s$.

  \begin{figure}
    \includegraphics{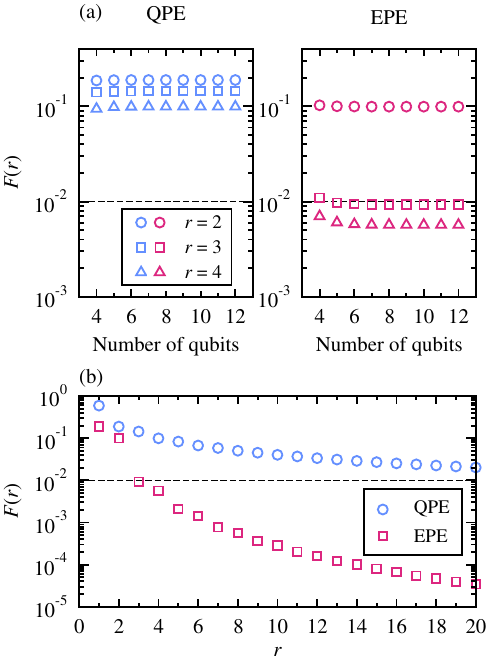}
    \caption{(a) $F(r)$ in Eq.~\eqref{eq:fr} as a function of $n_{q}$.
      (b) $F(r)$ as a function of $r$ for $n_{q} = 12$ ($N_{q} = 4096$).
    The results with different value of $n_{q}$ show a similar behavior.}
    \label{fig:qpeopt2}
  \end{figure}

  In order to find a reasonable value of $r$, in Fig.~\ref{fig:qpeopt2}(a)
  we plot the following complementary cumulative
  probability
  \begin{equation}
    F(r) = \max_{\theta} \bigl[1 - \sum_{i=1}^{r} P(k_{p_{i}}|\theta) \bigr]
    \label{eq:fr}
  \end{equation}
  for $r=2,3,4$ and $n_{q}=4, \ldots, 12$,
  where $\{k_{p_{i}}\}$ are permutations of $\{k\}$ such that $P(k_{p_{1}}|\theta) \geq P(k_{p_{2}}|\theta) \geq \dots$ for a given $\theta$.
  This quantity serves as a measure of leakage in QPE and EPE, and provides the confidence intervals of
  both approaches. In the QPE results  (Fig.~\ref{fig:qpeopt2}(a) left) with $r=4$, for example,
  $F(r)$ is approximately 0.1 for all $n_{q}$, which
  indicates that for each peak
  approximately 10\% of the measurement results will fall outside $r=4$ highest-probability points
  in the worst case.
  This leakage can make accurate estimation in the original QPE-based approaches challenging.
  This issue becomes particularly problematic when the isolated peak assumption is violated due to high density of states,
  and the tails of other peaks overlap the peak of interest.
  It can be seen from Fig.~\ref{fig:qpeopt2}(b) that simply increasing $r$ is not an efficient solution
  to this problem, as $F(r)$ in QPE decreases slowly with $r$.
  In contrast, the EPE results (Fig.~\ref{fig:qpeopt2}(a) right) exhibit
  a different behavior;
  one can see that the results with $r = 3$ fall below 0.01 for $n_{q} > 4$,
  which means that for each peak the measurement results will fall into one of these three points with a
  probability of more than 99\%.
  This indicates that, in EPE, for a single isolated peak well
  separated from others, the estimate of $|\bra{\psi_{s}}B\ket{\psi_{0}}|^2$ via Eq.~\eqref{eq:est_h} with $r=3$
  approaches to the true value within 1\% of error
  as the number of samples increases, and this can be a useful criterion when analyzing the results. 
  Also, Fig.~\ref{fig:qpeopt2}(b) shows that the value of $F(r)$ for EPE decreases much faster than the case of QPE; this can be interpreted as the quadratically fast decrease of the variance of the peak, or in other words the Heisenberg scaling of $F(r)$ with respect to $r$. 
  Based on this analysis, in our numerical simulations presented in the next section we use $r=3$.
  Choosing $r > 3$ improves the accuracy of the estimate in principle but may not be very practical,
  as one needs a large number of samples to get statistically reliable results.

  \section{numerical simulations}
  To get insights about the proposed approach, we perform numerical simulations of three simple systems.
  All our simulations are done with a noiseless statevector-based simulator in
  PennyLane library~\cite{bergholm2022pennylane}, therefore our results correspond to
  those in the limit of infinite sampling.
  The effect of finite sampling will be addressed in our future research.
  We employ the Trotter--Suzuki decomposition~\cite{Trotter,Suzuki1976},
  but other approximate time-propagation
  approaches~\cite{PhysRevLett.114.090502,7354428,PhysRevLett.118.010501,Low2019hamiltonian,PRXQuantum.2.040203} can
  also be used.

  \subsection{Spectral function in the electron-plasmon model}
  Our first example is a simple system describing one electron and
  plasmons in a single mode, which is used to model
  core electron photoemission in metals~\cite{lundqvist1067,PhysRevB.1.471,LarsHedin_1999}.
  The Hamiltonian is given as
  \begin{equation}
    H = \epsilon c^{\dagger} c + g c c^{\dagger} (b + b^{\dagger}) + \omega_{p} b^{\dagger} b,
  \end{equation}
  where $c$ and $c^{\dagger}$ are respectively creation and annihilation operators for the core electron of energy $\epsilon$,
  $b^{\dagger}$ and $b$ are bosonic operators for plasmons with energy $\omega_{p}$, and $g$ is the coupling coefficient.
  This model is exactly solvable~\cite{PhysRevB.1.471},
  and the explicit form of the
  hole part of the spectral function $A^{\textrm{hole}}(\omega)$ (Eq.~\eqref{eq:aw_hole}) is given as
  \begin{equation}
  A^{\textrm{hole}}(\omega) = \sum_{n=0}^{\infty} \frac{1}{n!}e^{-(g/\omega_{p})^{2}} \Bigl(\frac{g}{\omega_{p}}\Bigr)^{2n}
    \delta (\omega - \epsilon - \frac{g^{2}}{\omega_{p}} + n \omega_{p}).
    \label{eq:a_ep}
  \end{equation}
  Equation~\eqref{eq:a_ep} consists of multiple plasmon excitations whose peak weights
  decrease exponentially with respect
  to plasmon number $n$. 
  We calculate $A^{\textrm{hole}}(\omega)$ in this system using the proposed approach.
  The plasmons are treated using the method described in Ref.~\cite{PhysRevLett.121.110504}, which
  rewrites the bosonic part of the Hamiltonian to those of the harmonic oscillators.
  In the ground state of this system
  the core-electron level is fully occupied and no plasmon exists, therefore no ancilla qubit is needed.
  The plasmon wavefunction is expressed by the Gaussian function, which can be encoded with several
  approaches~\cite{grover2002creating,kitaev2009wavefunction}.
  The parameters used in the simulation are $\epsilon=-1,\omega_{p}=1,g=0.8,n_{s}=6,$ and $T=0.8$.
  The time propagation operator $\exp[i (H - E_{0}) T]$ is implemented
  with the first-order Trotter--Suzuki decomposition with the time step $\Delta T = 0.025$.

  \begin{figure}[ht]
    \centering
      \includegraphics{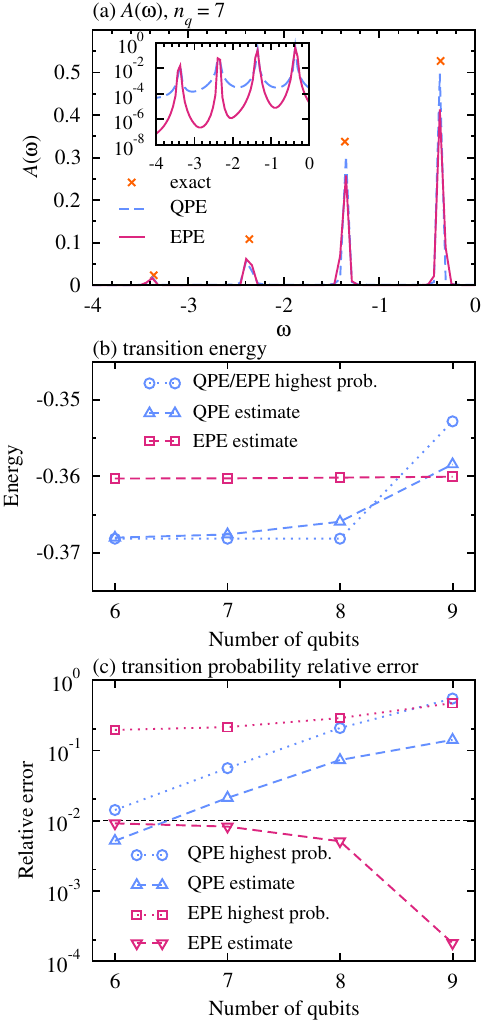}
      \caption{(a) $A^{\textrm{hole}}(\omega)$ of the electron-plasmon system calculated
      with QPE and EPE
      for $n_{q}=7$. The inset shows the same data in logarithmic scale.
        (b) The estimates of the transition energy corresponding to the first peak for $n_{q}=$ 6--9.
        (c) The relative estimation errors of the weight of the first peak for $n_{q}=$ 6--9.}
      \label{fig:eplasmon}
  \end{figure}

  Figure~\ref{fig:eplasmon}(a) shows $A^{\textrm{hole}}(\omega)$ calculated with QPE and
  EPE for $n_{q}=7$, together with the exact results from Eq.~\eqref{eq:a_ep}.
  In the plotted energy range there are four peaks whose heights decrease exponentially.
  The heights of EPE peaks are smaller than those in QPE, and as expected,
  the tail of the EPE spectrum has a much smaller intensity, as can be seen in the inset of the figure.
  Figure~\ref{fig:eplasmon}(b) shows the estimated transition energy corresponding to the first (i.e., $n=0$) peak
  via Eq.~\eqref{eq:est_e} calculated with QPE and EPE for $n_{q}=$ 6--9.
  For comparison, the frequency grid points $\omega_k$ with the highest probability in the QPE and EPE spectra
  are also shown.
  It can be seen that compared to the QPE estimates,
  which are more dependent on the values of the highest-probability frequency points,
  the EPE estimates
  are more robust and close to the exact result of $\epsilon + \frac{g^{2}}{\omega_{p}} = -0.36$ for all $n_{q}$.
  This result may be understood by noting that, as can be seen in Fig.~\ref{fig:qpeopt1}(b),
  for general $\theta$, the second largest probability value in the EPE probability distribution $(P(k|\theta))$
  is larger than its QPE counterpart.
  
  Figure~\ref{fig:eplasmon}(c) shows the relative errors of
  the estimated weights (transition probabilities) of the first peak
  from the exact analytic result of $e^{-(\frac{g}{\omega_{p}})^{2}}\approx 0.527$,
  obtained via Eq.~\eqref{eq:est_h} and also from the highest probability values extracted from the QPE and EPE spectra.
  Although the EPE peak heights (highest probability values) extracted from the spectra show large errors,
  as can also be seen in Fig.~\ref{fig:eplasmon}(a),
  EPE yields more accurate estimates than QPE.
  The relative errors in EPE are less than 0.01 for all $n_{q}$, which is consistent with our
  analysis for an isolated peak in the previous section.
  This result is reasonable because, as shown in the inset of Fig.~\ref{fig:eplasmon}(a),
  each of the EPE peaks has a very small overlap with neighboring peaks.
  The decrease in the EPE estimation error at $n_{q}=9$ originates from the fact
  that the value of the exact transition energy in this case is positioned near the midpoint between
  two frequency grid points.
  This situation is similar to the case in Fig.~\ref{fig:qpeopt1}(c),
  where the estimation scheme in Eq.~\eqref{eq:est_h} becomes exact for $r \geq 2$ in EPE.

  \subsection{Dipole transitions of the H$_{2}$O molecule}
    \begin{figure*}[htb]
      \centering
      \includegraphics{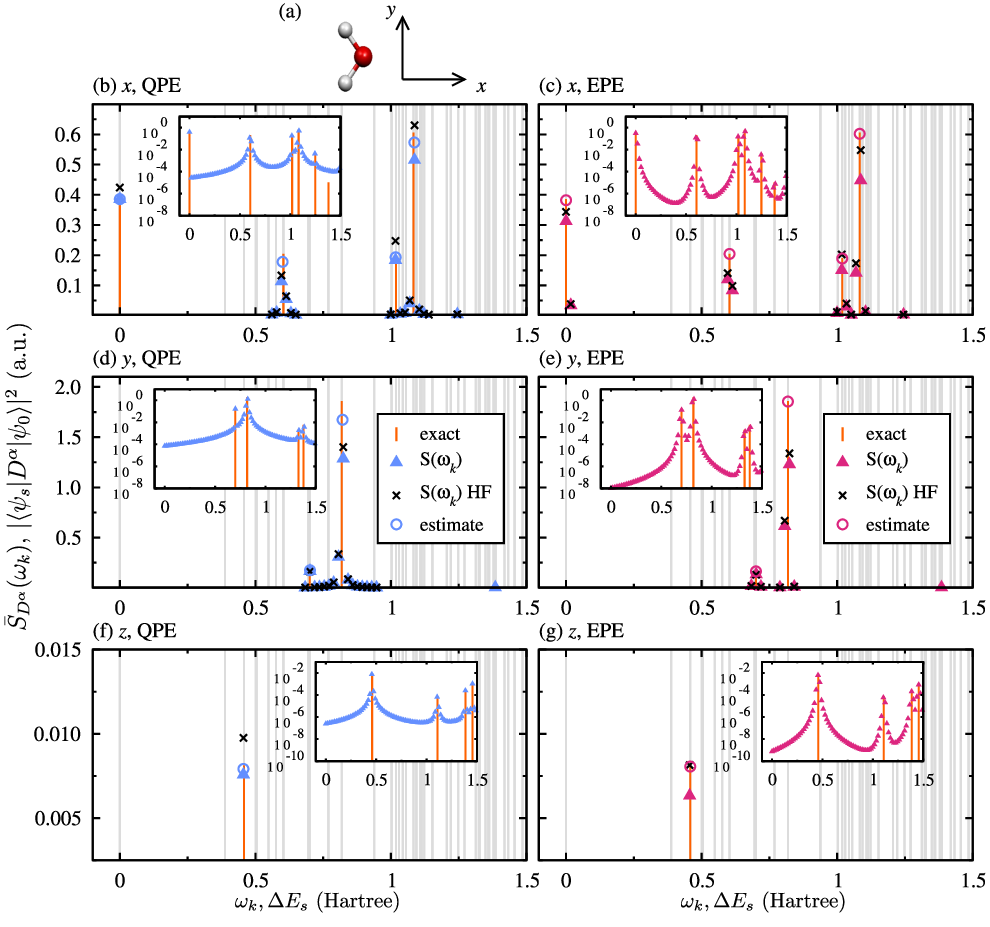}
      \caption{(Color online) (a) The orientation of the H$_{2}$O molecule in the simulation.
      (b--g) $\bar{S}_{D^{\alpha}}(\omega_{k})$ and the estimates of $\Delta E_{s}$
      and $|\bra{\psi_{s}}D^{\alpha}\ket{\psi_{0}}|^{2}$ via Eqs.~\eqref{eq:est_e} and \eqref{eq:est_h} for $\alpha=x,y,z$
      calculated with QPE and EPE.
       The exact spectra and the positions of
      all eigenstate levels are indicated by the vertical orange (dark gray) and light gray bars, respectively. The
      insets show the results in logarithmic scale.}
      \label{fig:h2o}
  \end{figure*}
    \begin{table*}[htb]
    \caption{The estimated values of $\Delta E_{s}$ and $|\bra{\psi_{s}}D^{\alpha}\ket{\psi_{0}}|^{2}$ ($\alpha=x,y,z$),
  and the relative estimation errors of $|\bra{\psi_{s}}D^{\alpha}\ket{\psi_{0}}|^{2}$  for selected transitions.}
    \begin{tabular*}{14cm}{@{\extracolsep{\fill}}lrrrrrrrr}
    \hline\hline
    & \multicolumn{3}{c}{$\Delta E_{s}$ (Hartree)} & \multicolumn{3}{c}{$|\bra{\psi_{s}}D^{\alpha}\ket{\psi_{0}}|^{2}$ (a.u.)}
    & \multicolumn{2}{c}{Relative error (\%)} \\
    $\alpha$ & Exact & QPE & EPE & Exact & QPE & EPE & QPE & EPE \\
    \hline
    $x$ & 0.000 & 0.000 & 0.000 & 0.386 & 0.386 & 0.382 & 0.018 & 0.933\\
        & 0.603 & 0.601 & 0.603 & 0.205 & 0.178 & 0.204 & 12.835 & 0.121\\
        & 1.019 & 1.017 & 1.019 & 0.190 & 0.193 & 0.189 & 1.700 & 0.577\\
        & 1.083 & 1.086 & 1.083 & 0.605 & 0.573 & 0.602 & 5.302 & 0.594\\
    $y$ & 0.700 & 0.701 & 0.700 & 0.166 & 0.175 & 0.165 & 5.521 & 0.970\\
        & 0.818 & 0.822 & 0.818 & 1.860 & 1.673 & 1.854 & 10.025 & 0.311\\
    $z$ & 0.458 & 0.456 & 0.458 & 0.008 & 0.008 & 0.008 & 2.487 & 0.795\\
    \hline\hline
  \end{tabular*}
  \label{tbl:h2o}
  \end{table*}
  Our next example is a quantum chemistry application of the current approach. We consider
  the dipole operators $D^{\alpha}$ as $B$ in Eq.~\eqref{eq:sbw}, where
  \begin{equation}
    D^{\alpha} = -\sum_{\mu\nu}d_{\mu\nu}^{\alpha} \bigl(c^{\dagger}_{\mu}c_{\nu} + c^{\dagger}_{\nu}c_{\mu}\bigr)
    + D^{c,\alpha} + D^{n,\alpha}.
    \label{eq:dipole_operator}
  \end{equation}
  Here $d_{\mu\nu}^{\alpha}$ are the dipole integrals defined as
  \begin{equation}
    d_{\mu\nu}^{\alpha} = \int \phi_{\mu}^{*}(\mathbf{r}) \alpha \phi_{\nu}(\mathbf{r}) d^{3}r,
  \end{equation}
  where $\alpha = x, y, z$ are the position operators, $\phi_{\mu}$ are
  the Hartree--Fock (HF) orbitals,
  and $D^{c,\alpha}, D^{n,\alpha}$
  are the contributions from the core electrons and the nuclei, respectively.
  The quantity $S_{D^{\alpha}}(\omega)$
  contains information about
  the transition dipole moments $|\bra{\psi_{s}}D^{\alpha}\ket{\psi_{0}}|^{2}$
  and the corresponding transition energies $\Delta E_{s}$,
  from which the photoabsorption cross section and the oscillator strengths of a molecule
  can be calculated~\cite{10.1093/acprof:oso/9780199563029.001.0001}.
  It is important to stress that in this approach
  no prior knowledge of the excited-state wavefunctions is required to calculate the excitation spectra.
  This is in contrast to approaches which calculate transition probabilities
  between some specific state pairs~\cite{Ibe_2022,https://doi.org/10.1002/qute.202300042}.

  We consider the H$_{2}$O molecule with the orientation illustrated in Fig.~\ref{fig:h2o}(a).
  We use the STO-3G basis set, and freeze the lowest two spatial HF orbitals. Therefore, our active space consists of
  5 spatial HF orbitals (10 HF spin orbitals) and 6 electrons.
  We encode the system with the JWT, and employ the qubit tapering technique~\cite{bravyi2017tapering,doi:10.1021/acs.jctc.0c00113},
  using the particle-number conservation for spin-up and spin-down electrons.
  The number of qubits for the system after tapering is 8.
  The Hamiltonian matrix elements and the dipole integrals are taken from
  PennyLane Quantum Chemistry Database~\cite{bergholm2022pennylane}. 
  The position operators in Eq.~\eqref{eq:dipole_operator} are encoded with the LCU approach
  as explained in the method section.
  The number of Pauli operators are 19, 16, and 8 for $\alpha = x, y,$ and $z$, respectively,
  requiring at most $5$ ancilla qubits. The parameters for QPE are $T=1.5$ a.u. and $n_{q}=8$, and
  the second-order Trotter--Suzuki decomposition is used with the time step $\Delta T = \frac{1.4}{32} = 0.04375$.

  Figures~\ref{fig:h2o}(b--g) show $\bar{S}_{D^{\alpha}}(\omega_{k})$ calculated with QPE and EPE
  and the estimates of the transition energies and the corresponding dipole moments using Eqs.~\eqref{eq:est_e} and \eqref{eq:est_h}.
  The exact spectra and the positions of all eigenstate levels ($\Delta E_{s}$) are also indicated in the figures.
  The results using the HF ground state as $\ket{\psi_{0}}$ are also shown for comparison.
  It can be seen that only the $\alpha = x$ case (Figs.~\ref{fig:h2o}(b) and (c))
  shows a nonzero peak at $\Delta E_{s} = 0$.
  This reflects the fact that the ground state of this molecule has a nonzero dipole moment element
  $\bra{\psi_{0}} D^{\alpha}\ket{\psi_{0}}$ only for $\alpha = x$,
  when the orientation shown in Fig.~\ref{fig:h2o}(a) is used.
  This result suggests that, when $\bra{\psi_{0}} B \ket{\psi_{0}}$ is
  known to have a nonzero value, as in the present case, one does not necessarily need to know the exact value of $E_{0}$
  when constructing the quantum circuit in Fig.~\ref{fig:qc_sbw} but can extract it from the position of the first peak of the calculated spectra.

  When the HF wavefuntion is used as $\ket{\psi_{0}}$, the calculated peak heights change slightly although their peak positions remain unchanged. 
  This is because the HF wavefunction is written as a linear combination of eigenfunctions
  $\ket{\psi_{\rm{HF}}} = \sum_u d_u \ket{\psi_u}$, and therefore $\mathcal{P}(k)$ given in Eq.~\eqref{eq:p_k} becomes
  $\mathcal{P}(k) = \mathcal{N} \sum_{s,u} |d_u|^2 |\bra{\psi_{s}} B \ket{\psi_{u}}|^{2} P(k|\Delta E_{s} T)$.
  Note that in the current case the overlap between the exact and the HF wavefunctions is
  $|\bra{\psi_{\rm{HF}}}\psi_{0}\rangle|^{2} \approx 0.9796$.
  Practically, since the HF ground state can easily be obtained and encoded,
  it may conveniently be used to get approximate energy spectra
  with a low computational cost, as long as the electron correlation in the system is not very strong. In the case where an accurate evaluation of the transition probability is desired, connecting sophisticated ground-state wavefunction preparation techniques to this approach is required.

  As in the previous example, compared to the QPE results,
  the tail part of the EPE spectra decrease more rapidly, and the EPE peak heights are generally smaller.
  It can also be seen that the estimated $|\bra{\psi_{s}} D^{\alpha}\ket{\psi_{0}}|^{2}$ in EPE are again in
  better agreement with the exact results. To see this quantitatively,
  in Table~\ref{tbl:h2o} we list the estimated values of $\Delta E_{s}$ and $|\bra{\psi_{s}} D^{\alpha}\ket{\psi_{0}}|^{2}$
  for selected transitions, together with the relative estimation errors of $|\bra{\psi_{s}} D^{\alpha}\ket{\psi_{0}}|^{2}$.
  As for $\Delta E_{s}$, both QPE and EPE estimates obtained via Eq.~\eqref{eq:est_e}
  are in good agreement with the exact results, but EPE slightly
  outperforms QPE. For $|\bra{\psi_{s}} D^{\alpha}\ket{\psi_{0}}|^{2}$,
  the estimation errors of the EPE results are again less than 0.01 for all the cases.
  This indicates that
  the effects from the tails of other peaks are not very strong, therefore the approximation used in
  Eqs.~\eqref{eq:est_e} and \eqref{eq:est_h} is valid.
  The accuracy of the QPE estimates of $|\bra{\psi_{s}} D^{\alpha}\ket{\psi_{0}}|^{2}$ is,
  on the other hand, dependent on the values of $\Delta E_{s}$.
  The QPE estimate for state $s$ becomes accurate
  when the corresponding $\Delta E_{s}$ happens to be on or very close to some QPE frequency grid point $\omega_{k}$.
  In our calculation the transition with $\Delta E_{s}=0$ for $\alpha=x$ corresponds to this case, where the estimation
  error is $0.018$\%.
  The corresponding EPE spectrum has a nonzero tail part, as can be seen in Fig.~\ref{fig:h2o}(c),
  which is consistent with the result in Fig.~\ref{fig:qpeopt1}(a).
  For all other cases, EPE gives better agreement than QPE.
  This result suggests that QPE and EPE have complementary characteristics, and
  in practice it is possible to utilize both algorithms in order to derive accurate estimates.

  \subsection{Electromagnetic transitions of the $^6$Li nucleus}
  Our final example is the shell-model calculation of electromagnetic transition probabilities
  in nuclear physics (see the textbooks, for example, Ref.~\cite{ModernNuclearPhysics}).
  We consider the transitions from the ground state to the excited states through
  the electric (E) and magnetic (M) multipole operators, which are denoted as $\mathcal{M}_{\textrm{E}}(l,m)$
  and $\mathcal{M}_{\textrm{M}}(l,m)$, respectively, with $l=0,1,2,\dots$ the orbital angular momentum
  and $m=-l,\dots,+l$ its $z$-component.
  Our interest here is to evaluate the reduced transition probability
  from the initial state ($i$) to the final state ($f$)
  defined as~\cite{ModernNuclearPhysics}
  \begin{align}
    \mathcal{B}_{\textrm{E/M}}(l; i \to f) = \sum_{m,M_{f}} \abs{\mel{I^{\pi_{f}}_{f,n_{f}}M_{f}}{\mathcal{M}_{\textrm{E/M}}(l,m)}{I^{\pi_{i}}_{i,n_{i}}M_{i}}}^2 ,
    \label{eq:rtp}
  \end{align}
  where the states are labeled by the total angular momentum $I$, the $z$-component of the total angular momentum $M$, the parity $\pi$, and the additional quantum number to specify the states $n$.
  Because of the rotational symmetry,
  the states with the same $I^{\pi}_{n}$ but with different $M$ are degenerate.
  The reduced transition probabilities are related to the transition probabilities for emission and absorption of one photon
  through Fermi's golden rule~\cite{ModernNuclearPhysics}.
  The transition operators $\mathcal{M}_{\textrm{E/M}}(l,m)$ are one-particle operators and
  written in the second quantization form as
  \begin{equation}
    \mathcal{M}_{\textrm{E/M}}(l,m) = \sum_{\mu,\nu} \mathcal{M}_{\textrm{E/M}}(l,m)_{\mu\nu} c^{\dagger}_{\mu} c_{\nu}.
    \label{eq:m_em_lm}
  \end{equation}
  From these operators we construct the following Hermitian operators
  \begin{equation}
    \tilde{\mathcal{M}}_{\textrm{E/M}}(l) = \sum_{\mu,\nu} \sum_{m=0}^{+l}
    \tilde{\mathcal{M}}_{\textrm{E/M}}(l,m)_{\mu\nu} c^{\dagger}_{\mu} c_{\nu},
    \label{eq:tilde_m_em_l}
  \end{equation}
  where
  \begin{equation}
  \begin{aligned}
      &\tilde{\mathcal{M}}_{\textrm{E/M}}(l,m)_{\mu\nu} \\
      &= \begin{cases} 
\mathcal{M}_{\textrm{E/M}}(l,m)_{\mu\nu} & \text{for } m=0, \\
\mathcal{M}_{\textrm{E/M}}(l,m)_{\mu\nu} + (-1)^{m} \mathcal{M}_{\textrm{E/M}}(l,-m)_{\mu\nu} 
 & \text{for } m>0,\\
\end{cases}
  \end{aligned}
  \end{equation}
  and use them as the $B$ operator in Eq.~\eqref{eq:sbw}.
  The reduced transition probabilities $\mathcal{B}_{\textrm{E/M}}(l;i\to f)$
  and the corresponding transition energies $\Delta E_{f} = E_{f} - E_{i}$ are obtained
  from $\bar{S}_{B}(\omega_{k})$ using our estimation scheme given in Eqs.~\eqref{eq:est_e} and \eqref{eq:est_h}.
  \begin{figure}[htb]
    \centering
      \includegraphics{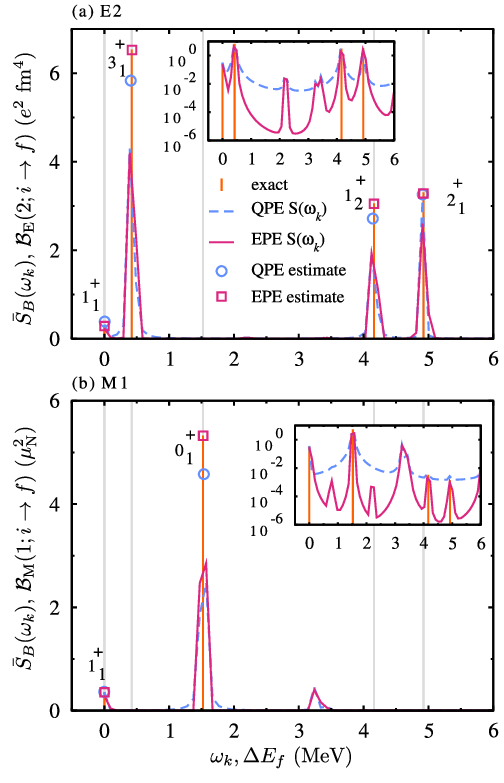}
      \caption{(Color online) $\bar{S}_{B}(\omega_{k})$ and the estimates of $\Delta E_{f}$ and $\mathcal{B}_{\textrm{E/M}}(l;i\to f)$
      calculated with QPE and EPE
      for (a) E2 and (b) M1.
      The insets show $\bar{S}_{B}(\omega_{k})$ in logarithmic scale. The exact reduced
      transition probabilities and the positions of all eigenstate levels are indicated by the orange
        (dark gray) and light gray bars,
      respectively.}
      \label{fig:shell}
  \end{figure}
  \begin{table*}
    \caption{The estimated values of $\Delta E_{f}$ and $\mathcal{B}_{\textrm{E/M}}(l;i\to f)$,
    and the relative estimation errors of $\mathcal{B}_{\textrm{E/M}}(l;i\to f)$.
    The transitions with $\Delta E_{f} = 0$ are related to the electric quadrupole and magnetic dipole moments of the ground state.
    The units of $\mathcal{B}_{\textrm{E/M}}(l;i\to f)$ are
    $e^{2}$fm$^{4}$ and $\mu_{\mathrm{N}}^{2}$ for E2 and M1 transitions, respectively.}
    \begin{tabular*}{14cm}{@{\extracolsep{\fill}}llrrrrrrrr}
    \hline\hline
    &                & \multicolumn{3}{c}{$\Delta E_{f}$ (MeV)}    &
    \multicolumn{3}{c}{$\mathcal{B}_{\textrm{E/M}}(l;i \to f)$} & \multicolumn{2}{c}{Relative error (\%)} \\
    Type & Transition & Exact & QPE & EPE & Exact & QPE & EPE & QPE & EPE \\
    \hline
    E2 & $1^{+}_{1} \to 1^{+}_{1}$ & 0.000 & 0.007 & 0.003 & 0.293 & 0.389 & 0.289 & 32.768 & 1.185\\
       & $1^{+}_{1} \to 3^{+}_{1}$ & 0.424 & 0.409 & 0.427 & 6.533 & 5.833 & 6.523 & 10.717 & 0.149\\
       & $1^{+}_{1} \to 1^{+}_{2}$ & 4.160 & 4.141 & 4.159 & 3.060 & 2.715 & 3.048 & 11.277 & 0.399\\
       & $1^{+}_{1} \to 2^{+}_{1}$ & 4.923 & 4.910 & 4.921 & 3.310 & 3.262 & 3.284 & 1.437 & 0.796\\
    M1 & $1^{+}_{1} \to 1^{+}_{1}$ & 0.000 & 0.000 & 0.004 & 0.356 & 0.366 & 0.354 & 2.769 & 0.461\\
       & $1^{+}_{1} \to 0^{+}_{1}$ & 1.523 & 1.536 & 1.525 & 5.325 & 4.579 & 5.324 & 14.022 & 0.022\\
    \hline\hline
  \end{tabular*}
  \label{tbl:shell}
  \end{table*}

   Here, we consider the $^6$Li nucleus in the conventional nuclear shell model, which has one neutron and one proton in the valence space of the $p$ shell consisting of 12 states (4 for the neutron and proton $p_{1/2}$ orbitals and 8 for the neutron and proton $p_{3/2}$ orbitals) 
  on top of the $^4$He closed core. We use the Cohen--Krath effective interaction~\cite{COHEN19651} for the model space of the $p$ shell.
  Recently a benchmark study of the variational quantum eigensolver and the
  unitary coupled-cluster calculations was reported for this nucleus~\cite{PhysRevC.106.034325}.
  The initial state is taken to be the ground state of the system, $I^{\pi_i}_{i,n_i} = 1^{+}_{1}$ with $M_{i}=0$.
  We note that the transitions between excited states are also of fundamental importance, and the current approach is applicable for these transitions in principle.
  As in the H$_2$O case above, we encode the system with the JWT, and apply
  the qubit tapering technique~\cite{bravyi2017tapering,doi:10.1021/acs.jctc.0c00113} using the particle number conservation of neutrons and protons, which reduces the number
  of qubits in the system register from 12 to 10.
  We consider the electric quadrupole ($\mathcal{M}_{\textrm{E}}(l=2)$) and the magnetic dipole ($\mathcal{M}_{\textrm{M}}(l=1)$) transitions, which are denoted
  as E2 and M1 transitions in short, respectively.
  The corresponding Hermitized transition operators, $\tilde{\mathcal{M}}_{\textrm{E}}(l=2)$
  and $\tilde{\mathcal{M}}_{\textrm{M}}(l=1)$,
  are expanded with 56 and 74 Pauli operators, respectively.
  Instead of applying LCU to these operators directly, which would require 6 ancilla qubits in both cases,
  we diagonalize the one-particle matrix elements of
  $\tilde{\mathcal{M}}_{\textrm{E/M}}(l)$ (Eq.~\eqref{eq:tilde_m_em_l})
  and express the Hamiltonian and the Hermitized transition operators
  in terms of the eigenvectors of these matrices.
  Since $\tilde{\mathcal{M}}_{\textrm{E/M}}(l)$ become diagonal in this basis,
  the number of Pauli operators is reduced to $12$ in both cases, reducing the number of LCU ancilla qubits from 6 to 4.
  The matrix elements of the nuclear Hamiltonian and transition operators are
  taken from the nuclear shell-model code KSHELL~\cite{SHIMIZU2019372,shimizu2013nuclear}.
  For the calculation of E2 transitions, the proton and neutron effective charges are set to 1.5$e$ and 0.5$e$, respectively.
  For the calculation of M1 transitions, the gyromagnetic ratio of orbital angular momentum is taken as 1.0 (0.0) for the proton (neutron),
  and that of spin is taken as 5.585 ($-3.826$) for the proton (neutron).
  The parameters for QPE are $T=1.0$ MeV$^{-1}$ and $n_{q} = 6$, and the second-order Trotter--Suzuki decomposition
  with the time step $\Delta T = \frac{1.0}{64}$ MeV$^{-1} = 0.015625$ MeV$^{-1}$ is used.

  Figures~\ref{fig:shell}(a) and (b) show the calculated $\bar{S}_{B}(\omega_{k})$ with QPE and EPE for E2 and M1 transitions.
  The estimates of $\Delta E_{f}$ and $\mathcal{B}_{\textrm{E/M}}(l;i\to f)$ via Eqs.~\eqref{eq:est_e} and \eqref{eq:est_h} are also shown
  together with the exact results.
  The estimated values and
  the relative estimation errors of $\mathcal{B}_{\textrm{E/M}}(l;i\to f)$
  are listed in Table~\ref{tbl:shell}.
  Because of the limited energy resolution in the current calculation due to the small value of $n_{q}$,
  the calculated peaks are broader compared to the H$_{2}$O case above.
  However, one can still observe the same trend as those observed in the previous two systems;
  the peaks in the EPE results are again more localized, and the estimated values of $\Delta E_{f}$
  and $\mathcal{B}_{\textrm{E/M}}(l;i \to f)$ in EPE 
  show
  a better agreement with the exact results.
  The relative estimation errors of $\mathcal{B}_{\textrm{E/M}}(l;i \to f)$ in EPE
  are again less than 1\% for most cases, as seen in Table~\ref{tbl:shell}.
  The one exception is the result at $\Delta E_{f} = 0$ for E2, whose relative error is slightly above 1\% (1.185\%).
  This error comes from the tail of the neighboring large peak in E2 at $\Delta E_{f} \approx 0.424$ MeV corresponding
  to $1^{+}_{1} \to 3^{+}_{1}$ transition, as seen in Fig.~\ref{fig:shell}(a).
  This tail also significantly worsens the QPE estimate of the first peak,
  whose relative error is as large as $\approx 33$\%, showing a clear advantage of EPE.
  Unlike the H$_{2}$O case above, the QPE estimate of the first peak at $\Delta E_{f} = 0$ in the M1 transition
  (Fig.~\ref{fig:shell}(b)) has a relatively large error of 2.769\%. This is also because of
  the leakage effect from the neighboring large peak corresponding to $1^{+}_{1} \to 0^{+}_{1}$ transition,
  even though these two peaks are separated by  $\approx 1.523$ MeV, corresponding
  to $\approx 16$ QPE frequency grid points in the current calculation. Overall, the current results again
  demonstrate the effectiveness of the EPE-based approach, especially when
  the number of QPE qubits ($n_{q}$) is limited, or when the peaks in the spectra are closely located to each other.

  Due to the high locality of the EPE probability distribution,
  in Figs.~\ref{fig:shell}(a) and (b) one can also observe some low-intensity peaks only in the EPE results.
  In the QPE spectra, these low-intensity peaks are obscured by the tails of other peaks due to the spectral leakage
  problem. This enhanced peak-detection capability is another potentially useful feature of EPE.

  We remark that some of the peaks in Figs.~\ref{fig:shell}(a) and (b) correspond to transitions to
  high-lying excited states, whose transition energies $\Delta E_{f}$
  are outside the range of $[0, \frac{2\pi}{T}]$.
  The most noticeable one is at around $3.2$ MeV in Fig.~\ref{fig:shell}(b).
  Practically, since the positions of these peaks
  depend on the value of $T$ as $\omega = E_{f} - E_{0} - \frac{2 \pi}{T} m_{f}$ with $m_{f}$ some positive integer,
  one can easily check if the observed peaks correspond to the transitions in the given energy range by varying $T$.
  For example, the peak at $\Delta E_{f} \approx 3.2$ MeV in Fig.~\ref{fig:shell}(b) 
  can be interpreted as the transition to the $2^+_3$ state in comparison with the exact value 
  ($\mathcal{B}_{\textrm{M}}(1;1^+_1 \to 2^+_3) \approx 0.562$ $\mu_{\mathrm{N}}^{2}$  
  at $\Delta E_{f} \approx 9.550$ MeV).

  \section{conclusions and outlook}
  In this work we have presented a quantum approach to calculate the spectral properties of many-body systems
  based on QPE with entangled input states, and have shown that the calculated spectra have remarkable properties.
  The most notable feature is that the peaks in the calculated energy spectra are much more localized than those obtained with the original QPE-based approaches, thanks to the entanglement-assisted decrease of uncertainty of the peak width.
  Taking advantage of this property, we have proposed a simple scheme to estimate both the transition energy and the corresponding transition probability
  of selected states from the calculated spectra.
  For the latter quantity, we have shown that the estimation error in the proposed scheme is guaranteed to be
  less than 1\% for isolated peaks in the limit of infinite sampling.
  Our proof-of-concept calculations show that the current approach is applicable to wide range of systems.

  Since both the entangled input states considered in this study (Eq.~\eqref{eq:a_j_qpe_opt}) and
  our estimation scheme can easily be implemented,
  they serve as a useful option for the QPE-based approaches.
  Other forms of the input states~\cite{patel2024optimal,greenaway2024case} as well as
  a more sophisticated estimation scheme should also be investigated.

  Future work will focus on practical aspects of this approach.
  While our simulations are based on an ideal statevector-based calculation with the exact
  ground state,
  in practice this approach requires accurate ground-state preparation and time-propagation schemes.
  In addition, as the number of samples (shots) available in real quantum devices is limited,
  it is important to develop techniques to sample only the most important energy region of the problem.
  Combining this approach with quantum amplitude amplification~\cite{gilles2002,10.1145/2591796.2591854}
  or recently proposed efficient
  filtering approaches~\cite{PRXQuantum.2.020321,PhysRevA.106.032420,zeng2022universal}
  could be an interesting research direction.

  \begin{acknowledgments}
    R.S. is very grateful to Kaito Wada for technical discussions about optimal phase estimation and informing him of Ref.~\cite{4655455}.
    R.S. also would like to thank Hiroshi Yano, Kohei Oshio, and Mika Yoshimoto for helpful discussions.
    This work was supported by Quantum Leap Flagship Program (Grant No. JPMXS0118067285 and No. JPMXS0120319794) from the MEXT, Japan, and the Center of Innovations for Sustainable Quantum AI (JPMJPF2221) from JST, Japan.
    K.S. acknowledges support from KAKENHI Scientific Research C (21K03407) and Transformative Research Area B (23H03819) from JSPS, Japan.
    T.A. acknowledges support from KAKENHI Scientific Research C (21K03564).
  \end{acknowledgments}

  \appendix*

  \section{A way to avoid controlled $\exp[iHT]$ operations and dependence on $E_{0}$}

\begin{figure}[htb]
    \centering
    \includegraphics[scale=0.62]{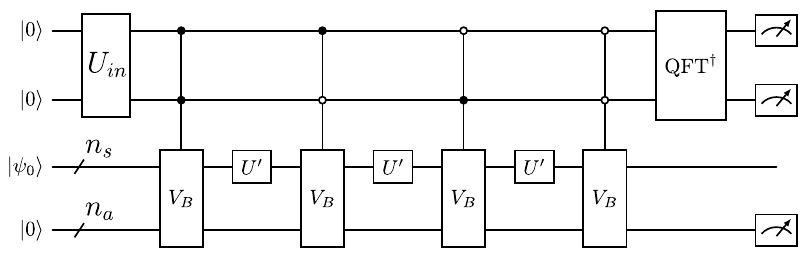}
    \caption{A quantum circuit without depending on $E_{0}$ and controlled $\exp[iHT]$ for $n_{q}=2$. Here $U'=\exp[i H T]$.}
    \label{fig:qpde}
  \end{figure}

  The quantum circuit shown in Fig.~\ref{fig:qc_sbw} depends on the ground-state energy $E_{0}$ and
  controlled $U=\exp [i(H - E_{0})T]$ operations.
  One possible way to remove these dependencies is to consider an operator whose action to
  a state $\ket{j}\ket{\psi_{0}}$ is given as (the ancilla register is omitted for simplicity)
  \begin{eqnarray}
    \ket{j}\ket{\psi_{0}} &\to& U'^{j} B U'^{N_{q} - 1 - j} \ket{j}\ket{\psi_{0}} \nonumber\\
    &=& e^{i E_{0} T(N_{q} - 1)} \nonumber\\
    & \times & \sum_{s} \bra{\psi_{s}} B \ket{\psi_{0}} e^{i (E_{s} - E_{0})T j} \ket{j} \ket{\psi_{s}},
    \label{eq:qpde}
  \end{eqnarray}
  where $U'=\exp \bigl [ i H T \bigr ]$ and $j=0,1,2,\dots,N_{q} - 1$.
  A quantum circuit implementing this idea
  is shown in Fig.~\ref{fig:qpde} for $n_{q} = 2$.
  The controlled $U$ operations are replaced with uncontrolled $U'$ that do not depend on $E_{0}$,
  and instead one needs $N_{q}$ multiply-controlled $V_{B}$ gates.
  This form can be advantageous when $V_{B}$ has a relatively simple form, as in the case of Fig.~\ref{fig:qc_ajw}.

  \bibliography{refs}

\end{document}